\documentclass[pra,twocolumn,amsmath,amssymb,superscriptaddress]{revtex4-1}

\usepackage{epsfig,amsmath}
\usepackage{subfigure}
\usepackage{graphicx}
\usepackage{dcolumn}
\usepackage{stmaryrd}
\usepackage{mathrsfs}
\usepackage{pifont}
\usepackage{amsthm}
\usepackage{amssymb}
\usepackage{bm}
\usepackage{latexsym}
\usepackage[colorlinks=true,linkcolor=blue,citecolor=blue]{hyperref}
\usepackage{color}
\usepackage{dsfont}
\usepackage{mathtools}

\theoremstyle{plain}

\newcommand{\lam}{\lambda}

\newcommand{\De}{\varDelta}

\newcommand{\ga}{\gamma}
\newcommand{\Ga}{\varGamma}
\newcommand{\om}{\omega}
\newcommand{\Om}{\varOmega}

\newcommand{\si}{\sigma}

\newcommand{\mC}{\mathcal{C}}
\newcommand{\mP}{\mathcal{P}}

\newcommand{\dif}{\mathrm{d}}
\newcommand{\e}{\mathrm{e}}

\begin{document}

\title{Trapping quantum coherence with a dissipative thermal bath}

\author{Jia-Ming Zhang}
\email{Email address: zhangjiaming@sdust.edu.cn}
\affiliation{College of Electronics and Information Engineering, Shandong University of Science and Technology, Qingdao 266590, Shandong, China}

\author{Bing Chen}
\affiliation{College of Electronics and Information Engineering, Shandong University of Science and Technology, Qingdao 266590, Shandong, China}

\author{Jun Jing}
\email{Email address: jingjun@zju.edu.cn}
\affiliation{School of Physics, Zhejiang University, Hangzhou 310027, Zhejiang, China}

\date{\today}

\begin{abstract}
In the long-time limit, an open quantum system coupled to a dissipative environment is believed to lose its coherence without driving or measurement. Counterintuitively, we provide a necessary condition on trapping the coherence of a two-level system entirely with a thermal bath. Based on a time-local master equation, it is found that the residue coherence survives even under a high-temperature bath as long as the long-time Lamb shift is exactly negative to the system transition frequency. This condition is generally met in the strong and even ultrastrong coupling regime that could be relaxed by increasing the environmental temperature. The counter-rotating interactions between system and bath is indispensable to the residue coherence, whose magnitude is affected by the system initial state and the bath structure.
\end{abstract}

\maketitle

{\em Introduction.---} One crucial task in quantum technology and engineering~\cite{Mandel1995,Vandersypen2005,Streltsov2017,Scully2012,Gardiner2004,Breuer2002} is to preserve the coherence of quantum systems when they are unavoidably coupled to surrounding environments. A number of strategies by coherent operations have been used to neutralize the environmental effect, including but not limited to quantum error correction~\cite{Shor1995}, dynamical decoupling~\cite{Viola1999}, quantum feedback control~\cite{Rabitz2000}, and coherent measurement~\cite{Cable2005,Filip2014}. Can environments themselves be used to hold the system coherence? The answer is positive. Environmental memory effect has been exploited to induce the long-time coherence in biological systems~\cite{Chin2013} and the steady-state entanglement in coherently coupled dimer systems~\cite{Huelga2012}. In non-Markovian environments, the system coherence can be hold by either vanishing decay rate in the long-time limit~\cite{Tong2010} or pure-dephasing dynamics~\cite{Addis2014}. These phenomena can be categorized as coherence trapping. The magnitude of the steady-state or residue coherence is affected by the spectral density and the initial system-bath correlation~\cite{Zhang2018a}. When taking a general scenario of dissipative open quantum system into consideration, system coherence under the weak-coupling regime disappears in the long-time limit~\cite{Leggett1987} mostly due to the noncommutativity between free Hamiltonian and interaction Hamiltonian. Guarnieri et al. indicated that, an interaction Hamiltonian that has nonzero projections over both parallel and orthogonal components with respect to the free system Hamiltonian constitutes a sufficient condition for holding a residue coherence of a generic two-level system coupled to thermal bath~\cite{Guarnieri2018,Roman2021}.

Here for a generic unbiased spin-boson model, we provide a necessary condition under which a thermal bath could become a unique resource to trap the coherence of a dissipative two-level system, even with a high temperature. This condition is about the environmental-induced Lamb shift, which is determined by the system transition frequency, the bath structure and temperature, and the system-bath coupling strength. It is found that our condition for residue coherence is generally satisfied in the ultrastrong coupling regime, that has been approached in many platforms. Moreover, we investigate the system evolution within two characteristic timescales and find that the magnitude of the steady-state coherence depends on the bath structure and the system initial state.

{\em Model and coherence-trapping condition.---} Consider a two-level system or spin-$1/2$ with transition frequency $\Om$ undergoing a dissipative dynamics in a thermal bosonic bath. The full Hamiltonian in the Schr\"{o}dinger picture ($\hbar\equiv 1$) reads,
\begin{equation}\label{H}
H=\frac{\Om}{2}\si_z +\sum_j\om_j a_j^\dag a_j +\lam\si_x\sum_j\left(g_ja_j+g_j^*a_j^\dag\right).
\end{equation}
Here $\si_z$ and $\si_x$ are the Pauli matrices. $j$ indexes various field modes of the bath with eigenfrequencies $\om_j$ and creation (annihilation) operators $a_j^\dag$ ($a_j$). $\lam$ is a dimensionless system-bath coupling constant that is factorized from the relative coupling strengths $g_j$ and $g_j^*$. The system-bath interaction Hamiltonian does not commute with the free Hamiltonian, which induces the dynamics of both system population and system coherence.

Applying the well-developed projection operator technique and the time-convolutionless expansion of the dynamical generator up to the second-order, we deduce the second-order time-local master equation for the two-level system~\cite{Breuer2002,Kofman2004}. The reduced density matrix of the system is linearly decomposed with the Pauli matrices and the identity matrix $\rho_{\rm S}=\frac{1}{2}(\mathds{1}_2+\bm{r}\cdot\bm{\si})$ to attain a non-Markovian Bloch equation. Here $\bm{r}$ is the Bloch vector with three real components $x, y, z={\rm Tr}_{\rm S}(\si_{x,y,z}\rho_{\rm S})$. In particular, $x$ and $y$ denote respectively the real and imaginary parts of the system coherence and $z$ the system population imbalance. Given the free system Hamiltonian~(\ref{H}) contains only $\si_z$, the system coherence that we focus on is decoupled from the system population imbalance. The coherence can be described by ${\bm r}_{xy}\equiv[x,y]^{\rm T}$, the projection of $\bm{r}$ onto the transverse plane (the $x-y$ plane), whose equation of motion is
\begin{equation}\label{xy}
\frac{\dif}{\dif t}\bm{r}_{xy}(t)=\bm{M}_{xy}(t)\bm{r}_{xy}(t),
\end{equation}
with
\begin{equation}\label{Mxy}
\bm{M}_{xy}(t)=
\begin{pmatrix}
0 & -\Om \\
\Om+\De(t) & -\Ga(t)
\end{pmatrix}.
\end{equation}
Here the time-dependent coefficients
\begin{equation}\label{coeff}
\begin{aligned}
\De(t)\coloneqq&\ 4\lam^2\int_0^t \dif t'\int_0^\infty \dif\om\ J_{\rm eff}(\om;T)\cos{\om t'}\sin{\Om t'},\\
\Ga(t)\coloneqq&\ 4\lam^2\int_0^t \dif t'\int_0^\infty \dif\om\ J_{\rm eff}(\om;T)\cos{\om t'}\cos{\Om t'},\\
\end{aligned}
\end{equation}
are the Lamb shift and the decoherence rate of the system, respectively. Both of them depend on the coupling constant $\lam$, the system transition frequency $\Om$, and the environmental temperature $T$. $J_{\rm eff}(\om;T)=J(\om)\coth(\om/2T)$ $(k_{\rm B}\equiv1)$ denotes the effective spectral density for a finite temperature $T$ with $J(\om)$ the vacuum-field (zero-temperature) spectrum. Detailed derivations for Eqs.~(\ref{xy}), (\ref{Mxy}), and (\ref{coeff}) are provided in Sec.~A of Supplementary Material (SM).

In most situations, the system coherence vector ${\bm r}_{xy}$ has a unique equilibrium point at $x=y=0$~\cite{Slotine1991}, as long as the coefficient matrix $\bm{M}_{xy}(t)$ in Eq.~(\ref{Mxy}) is nonsingular in the long-time limit. However, we find $\det\bm{M}_{xy}(\infty)=0$ under certain conditions. Particularly, the steady-state coherence survives with a nonvanishing $x(t)$ and a vanishing $y(t)$ for our Hamiltonian~(\ref{H}). To trap the coherence entirely by the thermal bath, it is required that 
\begin{equation}\label{condition}
\De(\infty)=\De(\infty; \lam, \Om, T)=-\Om,
\end{equation}
where
\begin{equation}\label{De}
\De(\infty;\lam,\Om,T)=-4\lam^2\Om\mP\int_0^\infty \dif\om \frac{J_{\rm eff}(\om;T)}{\om^2-\Om^2}
\end{equation}
is the long-time Lamb shift in Eq.~(\ref{coeff}). Here $\mP(\cdot)$ denotes the Cauchy principal integral.

Condition~(\ref{condition}) is a necessary criterion to have a nonvanishing steady-state coherence, which is the long-time solution of Eq.~(\ref{xy}) generally solved by numerical simulation. In the following, we assume that the vacuum-field spectrum $J(\omega)$ can be characterized by an Ohmic spectrum with a Lorentzian cutoff, i.e., $J(\omega)=J(\om;\ga,\om_0)=\om\ga^2/[(\om-\om_0)^2+\ga^2]$~\cite{Erez2008,Gordon2009,Zhang2019}. In the high-temperature limit, $J_{\rm eff}(\om;T)\rightarrow J_{\rm H}(\om;\ga,\om_0,T)=2T\ga^2/[(\om-\om_0)^2+\ga^2]$ due to the fact that $\lim_{T\to\infty}\coth(\om/2T)=2T/\om$, which describes a leaky cavity with not-so-high finesse mirrors~\cite{Scully2012}. Here $\om_0$ is the Lorentzian peak frequency and $\ga$ is the Lorentzian width whose reciprocal $\ga^{-1}$ characterises the bath memory time. In addition, if the Lorentzian peak frequency $\om_0\to 0$, this Ohmic spectrum reduces to the simplest form with Drude regularization, which usually describes the memory-friction damping of a quantum system~\cite{Breuer2001,Weiss2008}.

\begin{figure}[htbp]
\centering
\includegraphics[width=0.90\linewidth]{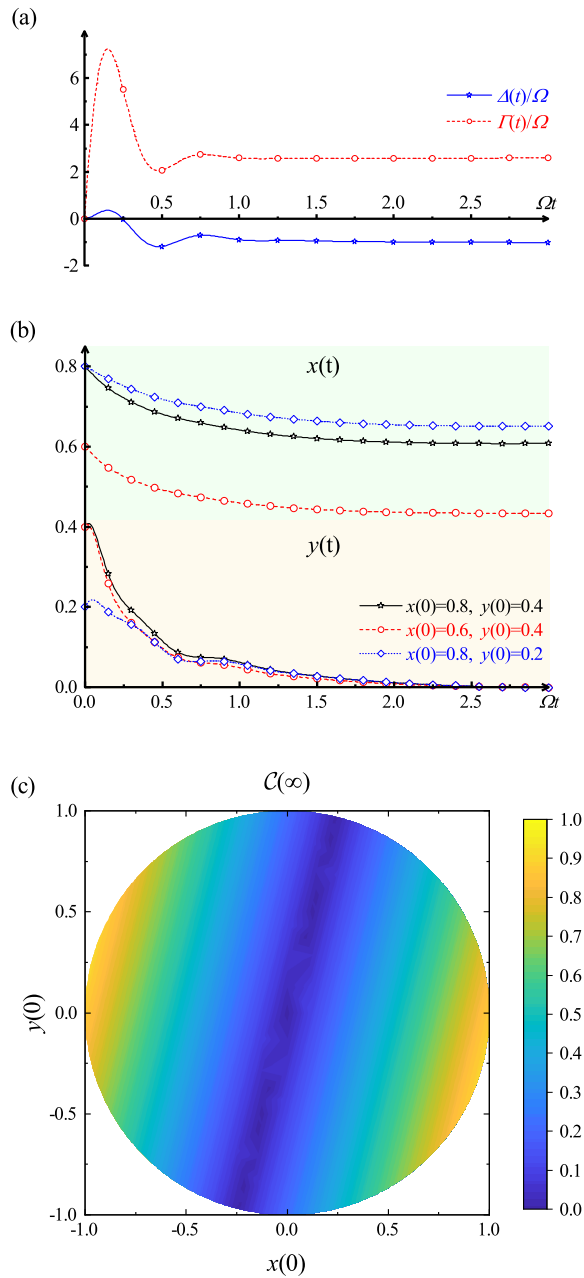}
\caption{(a) Dynamics of Lamb shift $\De(t)$ and decoherence rate $\Ga(t)$ for the dissipative two-level system. (b) Dynamics of the coherence components $x(t)$ (green area) and $y(t)$ (orange area) for various initial states: $x(0)=0.8$ and $y(0)=0.4$, $x(0)=0.6$ and $y(0)=0.4$, and $x(0)=0.8$ and $y(0)=0.2$. (c) Steady-state coherence $\mC(\infty)$ in the plane of $x(0)$ and $y(0)$. The system-bath coupling constant $\lam\approx0.09$ is determined by the condition~(\ref{condition}) under $T=100\Om$, $\ga=5\Om$, and $\om_0=10\Om$.}\label{Dxy0}
\end{figure}

{\em Residue coherence.---} To show the residue coherence $\mC=\sqrt{x^2+y^2}$~\cite{Baumgratz2014,Streltsov2017} could be survival in the presence of a high-temperature bath, here we choose $T=100\Om$. The memory coefficient and the Lorentzian peak frequency for the bath are set as $\ga=5\Om$ and $\om_0=10\Om$, respectively. Consequently, the system-bath coupling strength is found to be $\lam\approx0.09$ due to the condition~(\ref{condition}). In Figs.~\ref{Dxy0}(a) and (b), we plot the time evolutions of the system Lamb shift $\De(t)$ and the system decoherence rate $\Ga(t)$ based on Eq.~(\ref{coeff}) and those of the coherence components $x(t)$ and $y(t)$ starting from various initial states, respectively. All the dynamics can be roughly divided into two stages, characterized by the memory timescale of the bath $\tau_{\rm B}\sim\ga^{-1}$ and the stationary relaxation timescale of the system $\tau_{\rm S}\sim\Ga^{-1}(\infty)$, respectively. Note the latter can be obtained by the Fermi's golden rule $\Ga(\infty)=2\pi\lam^2 J_{\rm eff}(\Om)$.

In the first stage with a timescale $\tau_{\rm B}$, both $\De(t)$ and $\Ga(t)$ in Fig.~\ref{Dxy0}(a) experience a smooth fluctuation, that strongly depends on the bath structure. They start from zero and evolve to nonvanishing stable values in the end of this stage. $\Ga(\infty)\neq0$ rules out the possibility that the residue coherence results from a vanishing decay rate~\cite{Tong2010}. In the mean time, both components of the coherence vector $x(t)$ and $y(t)$ experience a comparatively fast decay. Their behaviors are determined by the two eigenvalues of matrix ${\bm M}_{xy}(t)$ in Eq.~(\ref{Mxy}). If both of them are real numbers, i.e., $\Ga^2(t)-4\Om[\Om+\De(t)]>0$, the quantities $x(t)$ and $y(t)$ then decay in a biexponential way as shown in Fig.~\ref{Dxy0}(b). If they are complex numbers, i.e., $\Ga^2(t)-4\Om[\Om+\De(t)]<0$, then they have certain oscillations in dynamics. More results about the Lamb shift under various parameters can be found in Sec.~B of SM.

Suppose that the dynamics of both components approaches the stage of stationary relaxation around a moment $t_0\sim\tau_{\rm B}$, after which $y(t)$ decays exponentially from $y(t_0)$ to zero with a decay rate denoted by $\Ga(\infty)\sim1/\tau_S$, i.e., $y(t)=y(t_0)\e^{-(t-t_0)/\tau_{\rm S}}$. Consequently, we have $x(t)=x(t_0)-\Om \left[y(t_0)-y(t)\right]\tau_{\rm S}$ by virtue of the Bloch equation~(\ref{xy}). Both $x(t)$ and $y(t)$ are reasonably dependent on the initial system state $x(0)$ and $y(0)$ for the evolution of the first stage, which significantly affects the subsequent stage. It can be illustrated by the evolutions in Fig.~\ref{Dxy0}(b) starting from various initial states under the same set of bath parameters. It is found that $|x(\infty)|$ increases with the magnitude of $x(0)$ for a fixed $y(0)$ and decreases with the magnitude of $y(0)$ for a fixed $x(0)$.

It is worth pointing out the coherence components $x(t)$ and $y(t)$ are symmetrical to each other when the interaction Hamiltonian that is proportional to $\si_x$ in Hamiltonian~(\ref{H}) is switched to $\si_y$. In that case, $x(t)$ will deplete and $y(t)$ could approach an non-zero value. This exchange does not change the residue coherence $\mC(\infty)$.

In the current model, the residue coherence can be obtained as $\mC(\infty)=|x(\infty)|=\left|x(t_0)-\Om y(t_0)\tau_{\rm S}\right|$, which establishes the relation with the system state at time $t_0$. Moreover, its dependence of the initial state ${\bm r}_{xy}(0)=[x(0), y(0)]^{\rm T}$ is displayed in Fig.~\ref{Dxy0}(c). In numerically calculation, the positivity of system density matrix is guaranteed, i.e., $\det\rho_{\rm S}=(1-|\bm{r}|^2)/4\geq 0$. Based on the Bloch equation~(\ref{xy}), the configuration in Fig.~\ref{Dxy0}(c) is anti-centrosymmetric with respect to the trivial incoherent state $x(0)=y(0)=0$. In another word, the steady-state coherence $\mC(\infty)$ corresponding to the initial state $-x(0)$ and $-y(0)$ has the same magnitude as that to the initial state $x(0)$ and $y(0)$. In addition, the configuration is symmetric about an axis across the center, whose slope is dependent on the parameters of the thermal bath (More results are provided in Sec.~B of SM). With certain initial states, one can obverse a finite system coherence is effeciently trapped in the long-time limit and the magnitude of $\mC(\infty)$ can be close to unit even under a high-temperature environment with $T=100\Om$.

\begin{figure}[htbp]
\centering
\includegraphics[width=0.95\linewidth]{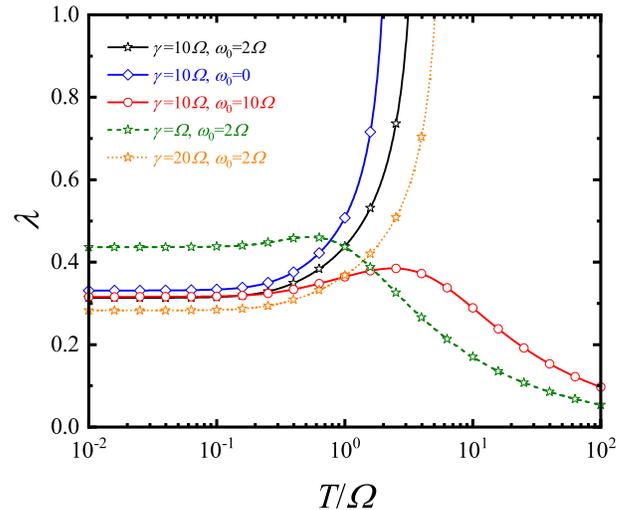}
\caption{Temperature dependence of the system-bath coupling constant $\lambda$ with various environmental memory coefficient $\gamma$ and spectral peak frequency $\omega_0$. All lines are numerically obtained by the coherence-trapping condition~(\ref{condition}). The black solid line, the blue solid line, and the orange dotted line represent the solutions that are divergent under certain high temperatures. The other two lines imply that the residue coherence can be survival even under high temperatures.}\label{lT}
\end{figure}

An associated question emerges from the perspective of the coherence-trapping condition~(\ref{condition}): in which case, can a Lamb shift as a function of temperature, system-environment coupling constant, and system transition frequency be negative to system transition frequency? Due to Eq.~(\ref{coeff}), the Lamb shift $\De$ is proportional to $\lambda^2$, meaning that the coupling constant $\lambda$ is lower bounded for a finite system transition frequency $\Om$. The facts that $\lim_{T\to\infty}\coth(\om/2T)=2T/\om$ and $\dif \coth[\om/(2T)]/\dif T ={\rm csch}^2[\om/(2T)]/(2T^2)>0$ indicate that the effective coupling strength associated with the spectrum of the thermal bath increases with the temperature. Therefore, $\lambda$ is also upper bounded for a high-temperature environment in order to satisfy the condition~(\ref{condition}). For a low- or moderate-temperature environment, the spectrum structure of the vacuum-field spectrum $J(\omega)$ would play a subtle role. The detailed analysis about various regimes of temperature are given in Sec.~C of SM.

Figure~\ref{lT} provides two typical situations with various spectral-parameters $\gamma$ and $\omega_0$, in which a proper coupling strength $\lam$ as a function of the temperature $T$ is evaluated by the condition~(\ref{condition}). It is interesting to found that if $\ga$ is much larger than $\om_0$ (see the blue, black, and orange lines), then no solution of $\lambda$ is found to satisfy the coherence-trapping condition for a sufficient high temperature. Otherwise, one can always find a proper $\lambda$ even when $T=100\Om$. In the latter case, $\lambda$ increases slowly with $T$ and over a moderate point, it becomes to decrease with $T$. Across four orders of $T/\Om$, $\lambda$ stays in the order of $0.1$, i.e., the strong and even ultrastrong coupling regime. It has been recently experimentally achieved in several platforms, such as superconducting quantum circuits, semiconductor quantum wells, and some hybrid quantum systems. To probe further see a recent review~\cite{Diaz2019}.

{\em Discussion.---} Some remarks are in order. The counter-rotating terms in the microscopic system-bath interaction Hamiltonian~(\ref{H}) are responsible for the virtual exchanges of energy between system and thermal bath. They are non-negligible in the ultrastrong coupling regime and indispensable to trap the steady-state coherence. Many physical systems, especially those in quantum optics, are described as $H^{\rm RWA}=\frac{1}{2}\Om\si_z +\sum_j\om_j a_j^\dag a_j +\lam\sum_j(g_j\si_+a_j+g_j^*\si_-a_j^\dag)$ under the rotating-wave approximation (RWA), where $\si_+$ and $\si_-$ are respectively the raising and lowering operators of the system. After a similar derivation using the second-order time-local master equation, we find that in the Bloch representation, the coherence and the population imbalance are still decoupled from each other. The dynamics of the coherence vector is similar to Eq.~(\ref{xy}) and the matrix $\bm{M}_{xy}(t)$ is modified as
\begin{equation*}
\bm{M}^{\rm RWA}_{xy}(t)=
\begin{pmatrix}
-\frac{1}{2}\Ga^{\rm RWA}(t) & -\Om-\frac{1}{2}\De^{\rm RWA}(t)\\
\Om+\frac{1}{2}\De^{\rm RWA}(t) & -\frac{1}{2}\Ga^{\rm RWA}(t)\\
\end{pmatrix},
\end{equation*}
where the Lamb shift and the decoherence rate under RWA can be found in Sec.~A of SM. It is important to notify that in the long-time dynamics, $\det\bm{M}^{\rm RWA}_{xy}(\infty)=[\Ga^{\rm RWA}(\infty)/2]^2+[\Om+\De^{\rm RWA}(\infty)/2]^2>0$. Therefore, in the absent of counter-rotating interactions, the coherence vector ${\bm r}_{xy}=[x(t),y(t)]^{\rm T}$ vanishes inevitably with a rate of $\Ga^{\rm RWA}(\infty)/2$ in the long-time scale. And the steady solution is an incoherent state.

Beyond RWA, various master equations have been proposed to describe the reduced dynamics of an open quantum system, such as the dynamical map approach~\cite{Rivas2017}, the partial-secular master equation~\cite{Jeske2015,Tscherbul2015}, and the coarse-grained master equation~\cite{Majenz2013,Davidovic2020,Mozgunov2020}. Most of them however cannot capture all the effects from the counter-rotating terms. For example, the coarse-grained master equation is formulated as a completely positive Lindblad-like equation, yet the effect of the counter-rotating terms is averaged out over a time-scale typical to the system motion, eventually leading to a vanishing coherence. It has been demonstrated that the second-order time-local master equation used in our work is significantly more accurate than those just mentioned. Its application range is even beyond the weak-coupling regime~\cite{Richard2020}. In addition, through incorporating more high-order perturbations to the time-local master equation, one can obtain the same form of the matrix $\bm{M}_{xy}$ of Eq.~(\ref{Mxy}) in the Bloch representation, except that the time-dependent coefficients $\De(t)$ and $\Ga(t)$ in Eq.~(\ref{coeff}) have a high-order correction~\cite{Breuer2001,Breuer2002}. The preceding analysis about the coherence trapping condition is not violated in quality.

The condition connecting the Lamb shift with the system energy spacing can be verified in platforms wherever the ultrastrong coupling is accessible. Moreover, it has been shown that a strong driving of a quantum system substantially enhances the Lamb shift induced by the environment~\cite{Gramich2014}, allowing it becomes comparable to the system transition frequency in magnitude. To experimentally trap the system coherence, one can measure the Lamb shift~\cite{Silveri2019} or use a qubit sensor to probe the environmental spectral density~\cite{Frey2017,Norris2018} and then rescale the coupling constant to satisfy the condition~(\ref{condition}). In our model, if the coupling constant can be manipulated at real time (e.g., the driven Dicke model~\cite{Bastidas2012} and the driven Jaynes-Cummings model~\cite{DeLiberato2009}) from the beginning of the dynamics, the coherence of the quantum system initialled in the $x-z$ plane [$y(0)=0$] of the Bloch sphere can be fully maintained.

One can also compare our results obtained by the time-local master equation with those by the equilibration theory~\cite{Mori2008}. It was strongly claimed that a quantum system coupled to a large-scale thermal bath should equilibrate with it, namely the system and the bath are in an entangled or correlated state. By the equilibration theory, the steady-state coherence cannot exist in a dissipative spin-boson model (A derivation is briefly given in Sec.~D of SM). However, the coherence trapping in our model and in an exactly solvable pure-dephasing spin-boson model~\cite{Addis2014} cannot be explained by the equilibration theory. It seems that whether or not the steady state of a quantum system coupled to a much larger thermal environment is a global Gibbs state is still an open question~\cite{Trushechkin2022}.

{\em Conclusion.---} We have provided a necessary condition for trapping (partial) coherence of the system in a general dissipative spin-boson model entirely with a thermal bath. It requires that the long-time Lamb shift is equivalent to the negative value of the system transition frequency. The coherence trapping is found to be induced by the counter-rotating interaction and strongly affected by the bath parameters and the system initial state. The condition discloses a fundamental consequence of ultrastrong coupling between system and bath.

We acknowledge grant support from the National Natural Science Foundation of China (Grants No. 11947079, No. 11974311, and
No. U1801661) and the Shandong Provincial Natural Science Foundation, China (Grants No. ZR2020QA079 and No. ZR2021MA081).

\bibliographystyle{apsrevlong}
\bibliography{SSC}

\begin{thebibliography}{43}%
\makeatletter
\providecommand \@ifxundefined [1]{%
 \@ifx{#1\undefined}
}%
\providecommand \@ifnum [1]{%
 \ifnum #1\expandafter \@firstoftwo
 \else \expandafter \@secondoftwo
 \fi
}%
\providecommand \@ifx [1]{%
 \ifx #1\expandafter \@firstoftwo
 \else \expandafter \@secondoftwo
 \fi
}%
\providecommand \natexlab [1]{#1}%
\providecommand \enquote  [1]{``#1''}%
\providecommand \bibnamefont  [1]{#1}%
\providecommand \bibfnamefont [1]{#1}%
\providecommand \citenamefont [1]{#1}%
\providecommand \href@noop [0]{\@secondoftwo}%
\providecommand \href [0]{\begingroup \@sanitize@url \@href}%
\providecommand \@href[1]{\@@startlink{#1}\@@href}%
\providecommand \@@href[1]{\endgroup#1\@@endlink}%
\providecommand \@sanitize@url [0]{\catcode `\\12\catcode `\$12\catcode
  `\&12\catcode `\#12\catcode `\^12\catcode `\_12\catcode `\%12\relax}%
\providecommand \@@startlink[1]{}%
\providecommand \@@endlink[0]{}%
\providecommand \url  [0]{\begingroup\@sanitize@url \@url }%
\providecommand \@url [1]{\endgroup\@href {#1}{\urlprefix }}%
\providecommand \urlprefix  [0]{URL }%
\providecommand \Eprint [0]{\href }%
\providecommand \doibase [0]{http://dx.doi.org/}%
\providecommand \selectlanguage [0]{\@gobble}%
\providecommand \bibinfo  [0]{\@secondoftwo}%
\providecommand \bibfield  [0]{\@secondoftwo}%
\providecommand \translation [1]{[#1]}%
\providecommand \BibitemOpen [0]{}%
\providecommand \bibitemStop [0]{}%
\providecommand \bibitemNoStop [0]{.\EOS\space}%
\providecommand \EOS [0]{\spacefactor3000\relax}%
\providecommand \BibitemShut  [1]{\csname bibitem#1\endcsname}%
\let\auto@bib@innerbib\@empty
\bibitem [{\citenamefont {Mandel}\ and\ \citenamefont
  {Wolf}(1995)}]{Mandel1995}%
  \BibitemOpen
  \bibfield  {author} {\bibinfo {author} {\bibfnamefont {L.}~\bibnamefont
  {Mandel}}\ and\ \bibinfo {author} {\bibfnamefont {E.}~\bibnamefont {Wolf}},\
  }\href@noop {} {\emph {\bibinfo {title} {Optical Coherence and Quantum
  Optics}}}\ (\bibinfo  {publisher} {Cambridge University Press},\ \bibinfo
  {year} {1995})\BibitemShut {NoStop}%
\bibitem [{\citenamefont {Vandersypen}\ and\ \citenamefont
  {Chuang}(2005)}]{Vandersypen2005}%
  \BibitemOpen
  \bibfield  {author} {\bibinfo {author} {\bibfnamefont {L.~M.~K.}\
  \bibnamefont {Vandersypen}}\ and\ \bibinfo {author} {\bibfnamefont {I.~L.}\
  \bibnamefont {Chuang}},\ }\bibfield  {title} {\bibinfo {title} {Nmr
  techniques for quantum control and computation},\ }\href {\doibase
  10.1103/RevModPhys.76.1037} {\bibfield  {journal} {\bibinfo  {journal} {Rev.
  Mod. Phys.}\ }\textbf {\bibinfo {volume} {76}},\ \bibinfo {pages} {1037}
  (\bibinfo {year} {2005})}\BibitemShut {NoStop}%
\bibitem [{\citenamefont {Streltsov}\ \emph {et~al.}(2017)\citenamefont
  {Streltsov}, \citenamefont {Adesso},\ and\ \citenamefont
  {Plenio}}]{Streltsov2017}%
  \BibitemOpen
  \bibfield  {author} {\bibinfo {author} {\bibfnamefont {A.}~\bibnamefont
  {Streltsov}}, \bibinfo {author} {\bibfnamefont {G.}~\bibnamefont {Adesso}}, \
  and\ \bibinfo {author} {\bibfnamefont {M.~B.}\ \bibnamefont {Plenio}},\
  }\bibfield  {title} {\bibinfo {title} {Colloquium: Quantum coherence as a
  resource},\ }\href {\doibase 10.1103/RevModPhys.89.041003} {\bibfield
  {journal} {\bibinfo  {journal} {Rev. Mod. Phys.}\ }\textbf {\bibinfo {volume}
  {89}},\ \bibinfo {pages} {041003} (\bibinfo {year} {2017})}\BibitemShut
  {NoStop}%
\bibitem [{\citenamefont {Scully}\ and\ \citenamefont
  {Zubairy}(2012)}]{Scully2012}%
  \BibitemOpen
  \bibfield  {author} {\bibinfo {author} {\bibfnamefont {M.~O.}\ \bibnamefont
  {Scully}}\ and\ \bibinfo {author} {\bibfnamefont {M.~S.}\ \bibnamefont
  {Zubairy}},\ }\href@noop {} {\emph {\bibinfo {title} {Quantum Optics}}}\
  (\bibinfo  {publisher} {Cambridge University Press},\ \bibinfo {address}
  {England},\ \bibinfo {year} {2012})\BibitemShut {NoStop}%
\bibitem [{\citenamefont {Gardiner}\ and\ \citenamefont
  {Zoller}(2004)}]{Gardiner2004}%
  \BibitemOpen
  \bibfield  {author} {\bibinfo {author} {\bibfnamefont {C.}~\bibnamefont
  {Gardiner}}\ and\ \bibinfo {author} {\bibfnamefont {P.}~\bibnamefont
  {Zoller}},\ }\href@noop {} {\emph {\bibinfo {title} {Quantum Noise}}}\
  (\bibinfo  {publisher} {Springer-Verlag},\ \bibinfo {address} {Berlin},\
  \bibinfo {year} {2004})\BibitemShut {NoStop}%
\bibitem [{\citenamefont {Breuer}\ and\ \citenamefont
  {Petruccione}(2002)}]{Breuer2002}%
  \BibitemOpen
  \bibfield  {author} {\bibinfo {author} {\bibfnamefont {H.-P.}\ \bibnamefont
  {Breuer}}\ and\ \bibinfo {author} {\bibfnamefont {F.}~\bibnamefont
  {Petruccione}},\ }\href@noop {} {\emph {\bibinfo {title} {The Theory of Open
  Quantum Systems}}}\ (\bibinfo  {publisher} {Oxford University Press},\
  \bibinfo {address} {New York},\ \bibinfo {year} {2002})\BibitemShut {NoStop}%
\bibitem [{\citenamefont {Shor}(1995)}]{Shor1995}%
  \BibitemOpen
  \bibfield  {author} {\bibinfo {author} {\bibfnamefont {P.~W.}\ \bibnamefont
  {Shor}},\ }\bibfield  {title} {\bibinfo {title} {Scheme for reducing
  decoherence in quantum computer memory},\ }\href {\doibase
  10.1103/PhysRevA.52.R2493} {\bibfield  {journal} {\bibinfo  {journal} {Phys.
  Rev. A}\ }\textbf {\bibinfo {volume} {52}},\ \bibinfo {pages} {R2493}
  (\bibinfo {year} {1995})}\BibitemShut {NoStop}%
\bibitem [{\citenamefont {Viola}\ \emph {et~al.}(1999)\citenamefont {Viola},
  \citenamefont {Knill},\ and\ \citenamefont {Lloyd}}]{Viola1999}%
  \BibitemOpen
  \bibfield  {author} {\bibinfo {author} {\bibfnamefont {L.}~\bibnamefont
  {Viola}}, \bibinfo {author} {\bibfnamefont {E.}~\bibnamefont {Knill}}, \ and\
  \bibinfo {author} {\bibfnamefont {S.}~\bibnamefont {Lloyd}},\ }\bibfield
  {title} {\bibinfo {title} {Dynamical decoupling of open quantum systems},\
  }\href {\doibase 10.1103/PhysRevLett.82.2417} {\bibfield  {journal} {\bibinfo
   {journal} {Phys. Rev. Lett.}\ }\textbf {\bibinfo {volume} {82}},\ \bibinfo
  {pages} {2417} (\bibinfo {year} {1999})}\BibitemShut {NoStop}%
\bibitem [{\citenamefont {Rabitz}\ \emph {et~al.}(2000)\citenamefont {Rabitz},
  \citenamefont {de~Vivie-Riedle}, \citenamefont {Motzkus},\ and\ \citenamefont
  {Kompa}}]{Rabitz2000}%
  \BibitemOpen
  \bibfield  {author} {\bibinfo {author} {\bibfnamefont {H.}~\bibnamefont
  {Rabitz}}, \bibinfo {author} {\bibfnamefont {R.}~\bibnamefont
  {de~Vivie-Riedle}}, \bibinfo {author} {\bibfnamefont {M.}~\bibnamefont
  {Motzkus}}, \ and\ \bibinfo {author} {\bibfnamefont {K.}~\bibnamefont
  {Kompa}},\ }\bibfield  {title} {\bibinfo {title} {Whither the future of
  controlling quantum phenomena?}\ }\href {\doibase
  10.1126/science.288.5467.824} {\bibfield  {journal} {\bibinfo  {journal}
  {Science}\ }\textbf {\bibinfo {volume} {288}},\ \bibinfo {pages} {824}
  (\bibinfo {year} {2000})}\BibitemShut {NoStop}%
\bibitem [{\citenamefont {Cable}\ \emph {et~al.}(2005)\citenamefont {Cable},
  \citenamefont {Knight},\ and\ \citenamefont {Rudolph}}]{Cable2005}%
  \BibitemOpen
  \bibfield  {author} {\bibinfo {author} {\bibfnamefont {H.}~\bibnamefont
  {Cable}}, \bibinfo {author} {\bibfnamefont {P.~L.}\ \bibnamefont {Knight}}, \
  and\ \bibinfo {author} {\bibfnamefont {T.}~\bibnamefont {Rudolph}},\
  }\bibfield  {title} {\bibinfo {title} {Measurement-induced localization of
  relative degrees of freedom},\ }\href {\doibase 10.1103/PhysRevA.71.042107}
  {\bibfield  {journal} {\bibinfo  {journal} {Phys. Rev. A}\ }\textbf {\bibinfo
  {volume} {71}},\ \bibinfo {pages} {042107} (\bibinfo {year}
  {2005})}\BibitemShut {NoStop}%
\bibitem [{\citenamefont {Filip}\ and\ \citenamefont
  {Marek}(2014)}]{Filip2014}%
  \BibitemOpen
  \bibfield  {author} {\bibinfo {author} {\bibfnamefont {R.}~\bibnamefont
  {Filip}}\ and\ \bibinfo {author} {\bibfnamefont {P.}~\bibnamefont {Marek}},\
  }\bibfield  {title} {\bibinfo {title} {Thermally induced creation of quantum
  coherence},\ }\href {\doibase 10.1103/PhysRevA.90.063820} {\bibfield
  {journal} {\bibinfo  {journal} {Phys. Rev. A}\ }\textbf {\bibinfo {volume}
  {90}},\ \bibinfo {pages} {063820} (\bibinfo {year} {2014})}\BibitemShut
  {NoStop}%
\bibitem [{\citenamefont {Chin}\ \emph {et~al.}(2013)\citenamefont {Chin},
  \citenamefont {Prior}, \citenamefont {Rosenbach}, \citenamefont
  {Caycedo-Soler}, \citenamefont {Huelga},\ and\ \citenamefont
  {Plenio}}]{Chin2013}%
  \BibitemOpen
  \bibfield  {author} {\bibinfo {author} {\bibfnamefont {A.~W.}\ \bibnamefont
  {Chin}}, \bibinfo {author} {\bibfnamefont {J.}~\bibnamefont {Prior}},
  \bibinfo {author} {\bibfnamefont {R.}~\bibnamefont {Rosenbach}}, \bibinfo
  {author} {\bibfnamefont {F.}~\bibnamefont {Caycedo-Soler}}, \bibinfo {author}
  {\bibfnamefont {S.~F.}\ \bibnamefont {Huelga}}, \ and\ \bibinfo {author}
  {\bibfnamefont {M.~B.}\ \bibnamefont {Plenio}},\ }\bibfield  {title}
  {\bibinfo {title} {The role of non-equilibrium vibrational structures in
  electronic coherence and recoherence in pigment–protein complexes},\ }\href
  {\doibase 10.1038/nphys2515} {\bibfield  {journal} {\bibinfo  {journal} {Nat.
  Phys.}\ }\textbf {\bibinfo {volume} {9}},\ \bibinfo {pages} {113} (\bibinfo
  {year} {2013})}\BibitemShut {NoStop}%
\bibitem [{\citenamefont {Huelga}\ \emph {et~al.}(2012)\citenamefont {Huelga},
  \citenamefont {Rivas},\ and\ \citenamefont {Plenio}}]{Huelga2012}%
  \BibitemOpen
  \bibfield  {author} {\bibinfo {author} {\bibfnamefont {S.~F.}\ \bibnamefont
  {Huelga}}, \bibinfo {author} {\bibfnamefont {A.}~\bibnamefont {Rivas}}, \
  and\ \bibinfo {author} {\bibfnamefont {M.~B.}\ \bibnamefont {Plenio}},\
  }\bibfield  {title} {\bibinfo {title} {Non-{Markovianity}-assisted steady
  state entanglement},\ }\href {\doibase 10.1103/PhysRevLett.108.160402}
  {\bibfield  {journal} {\bibinfo  {journal} {Phys. Rev. Lett.}\ }\textbf
  {\bibinfo {volume} {108}},\ \bibinfo {pages} {160402} (\bibinfo {year}
  {2012})}\BibitemShut {NoStop}%
\bibitem [{\citenamefont {Tong}\ \emph {et~al.}(2010)\citenamefont {Tong},
  \citenamefont {An}, \citenamefont {Luo},\ and\ \citenamefont
  {Oh}}]{Tong2010}%
  \BibitemOpen
  \bibfield  {author} {\bibinfo {author} {\bibfnamefont {Q.-J.}\ \bibnamefont
  {Tong}}, \bibinfo {author} {\bibfnamefont {J.-H.}\ \bibnamefont {An}},
  \bibinfo {author} {\bibfnamefont {H.-G.}\ \bibnamefont {Luo}}, \ and\
  \bibinfo {author} {\bibfnamefont {C.~H.}\ \bibnamefont {Oh}},\ }\bibfield
  {title} {\bibinfo {title} {Decoherence suppression of a dissipative qubit by
  the non-{Markovian} effect},\ }\href {\doibase
  10.1088/0953-4075/43/15/155501} {\bibfield  {journal} {\bibinfo  {journal}
  {J. Phys. B: At. Mol. Opt. Phys.}\ }\textbf {\bibinfo {volume} {43}},\
  \bibinfo {pages} {155501} (\bibinfo {year} {2010})}\BibitemShut {NoStop}%
\bibitem [{\citenamefont {Addis}\ \emph {et~al.}(2014)\citenamefont {Addis},
  \citenamefont {Brebner}, \citenamefont {Haikka},\ and\ \citenamefont
  {Maniscalco}}]{Addis2014}%
  \BibitemOpen
  \bibfield  {author} {\bibinfo {author} {\bibfnamefont {C.}~\bibnamefont
  {Addis}}, \bibinfo {author} {\bibfnamefont {G.}~\bibnamefont {Brebner}},
  \bibinfo {author} {\bibfnamefont {P.}~\bibnamefont {Haikka}}, \ and\ \bibinfo
  {author} {\bibfnamefont {S.}~\bibnamefont {Maniscalco}},\ }\bibfield  {title}
  {\bibinfo {title} {Coherence trapping and information backflow in dephasing
  qubits},\ }\href {\doibase 10.1103/PhysRevA.89.024101} {\bibfield  {journal}
  {\bibinfo  {journal} {Phys. Rev. A}\ }\textbf {\bibinfo {volume} {89}},\
  \bibinfo {pages} {024101} (\bibinfo {year} {2014})}\BibitemShut {NoStop}%
\bibitem [{\citenamefont {Zhang}\ \emph {et~al.}(2015)\citenamefont {Zhang},
  \citenamefont {Han}, \citenamefont {Xia}, \citenamefont {Yu},\ and\
  \citenamefont {Fan}}]{Zhang2018a}%
  \BibitemOpen
  \bibfield  {author} {\bibinfo {author} {\bibfnamefont {Y.-J.}\ \bibnamefont
  {Zhang}}, \bibinfo {author} {\bibfnamefont {W.}~\bibnamefont {Han}}, \bibinfo
  {author} {\bibfnamefont {Y.-J.}\ \bibnamefont {Xia}}, \bibinfo {author}
  {\bibfnamefont {Y.-M.}\ \bibnamefont {Yu}}, \ and\ \bibinfo {author}
  {\bibfnamefont {H.}~\bibnamefont {Fan}},\ }\bibfield  {title} {\bibinfo
  {title} {Role of initial system-bath correlation on coherence trapping},\
  }\href {\doibase 10.1038/srep13359} {\bibfield  {journal} {\bibinfo
  {journal} {Sci. Rep.}\ }\textbf {\bibinfo {volume} {5}},\ \bibinfo {pages}
  {13359} (\bibinfo {year} {2015})}\BibitemShut {NoStop}%
\bibitem [{\citenamefont {Leggett}\ \emph {et~al.}(1987)\citenamefont
  {Leggett}, \citenamefont {Chakravarty}, \citenamefont {Dorsey}, \citenamefont
  {Fisher}, \citenamefont {Garg},\ and\ \citenamefont {Zwerger}}]{Leggett1987}%
  \BibitemOpen
  \bibfield  {author} {\bibinfo {author} {\bibfnamefont {A.~J.}\ \bibnamefont
  {Leggett}}, \bibinfo {author} {\bibfnamefont {S.}~\bibnamefont
  {Chakravarty}}, \bibinfo {author} {\bibfnamefont {A.~T.}\ \bibnamefont
  {Dorsey}}, \bibinfo {author} {\bibfnamefont {M.~P.~A.}\ \bibnamefont
  {Fisher}}, \bibinfo {author} {\bibfnamefont {A.}~\bibnamefont {Garg}}, \ and\
  \bibinfo {author} {\bibfnamefont {W.}~\bibnamefont {Zwerger}},\ }\bibfield
  {title} {\bibinfo {title} {Dynamics of the dissipative two-state system},\
  }\href {\doibase 10.1103/RevModPhys.59.1} {\bibfield  {journal} {\bibinfo
  {journal} {Rev. Mod. Phys.}\ }\textbf {\bibinfo {volume} {59}},\ \bibinfo
  {pages} {1} (\bibinfo {year} {1987})}\BibitemShut {NoStop}%
\bibitem [{\citenamefont {Guarnieri}\ \emph {et~al.}(2018)\citenamefont
  {Guarnieri}, \citenamefont {Kol\'a\ifmmode~\check{r}\else \v{r}\fi{}},\ and\
  \citenamefont {Filip}}]{Guarnieri2018}%
  \BibitemOpen
  \bibfield  {author} {\bibinfo {author} {\bibfnamefont {G.}~\bibnamefont
  {Guarnieri}}, \bibinfo {author} {\bibfnamefont {M.}~\bibnamefont
  {Kol\'a\ifmmode~\check{r}\else \v{r}\fi{}}}, \ and\ \bibinfo {author}
  {\bibfnamefont {R.}~\bibnamefont {Filip}},\ }\bibfield  {title} {\bibinfo
  {title} {Steady-state coherences by composite system-bath interactions},\
  }\href {\doibase 10.1103/PhysRevLett.121.070401} {\bibfield  {journal}
  {\bibinfo  {journal} {Phys. Rev. Lett.}\ }\textbf {\bibinfo {volume} {121}},\
  \bibinfo {pages} {070401} (\bibinfo {year} {2018})}\BibitemShut {NoStop}%
\bibitem [{\citenamefont {Rom\'an-Ancheyta}\ \emph {et~al.}(2021)\citenamefont
  {Rom\'an-Ancheyta}, \citenamefont {Kol\'a\ifmmode~\check{r}\else \v{r}\fi{}},
  \citenamefont {Guarnieri},\ and\ \citenamefont {Filip}}]{Roman2021}%
  \BibitemOpen
  \bibfield  {author} {\bibinfo {author} {\bibfnamefont {R.}~\bibnamefont
  {Rom\'an-Ancheyta}}, \bibinfo {author} {\bibfnamefont {M.}~\bibnamefont
  {Kol\'a\ifmmode~\check{r}\else \v{r}\fi{}}}, \bibinfo {author} {\bibfnamefont
  {G.}~\bibnamefont {Guarnieri}}, \ and\ \bibinfo {author} {\bibfnamefont
  {R.}~\bibnamefont {Filip}},\ }\bibfield  {title} {\bibinfo {title} {Enhanced
  steady-state coherence via repeated system-bath interactions},\ }\href
  {\doibase 10.1103/PhysRevA.104.062209} {\bibfield  {journal} {\bibinfo
  {journal} {Phys. Rev. A}\ }\textbf {\bibinfo {volume} {104}},\ \bibinfo
  {pages} {062209} (\bibinfo {year} {2021})}\BibitemShut {NoStop}%
\bibitem [{\citenamefont {Kofman}\ and\ \citenamefont
  {Kurizki}(2004)}]{Kofman2004}%
  \BibitemOpen
  \bibfield  {author} {\bibinfo {author} {\bibfnamefont {A.~G.}\ \bibnamefont
  {Kofman}}\ and\ \bibinfo {author} {\bibfnamefont {G.}~\bibnamefont
  {Kurizki}},\ }\bibfield  {title} {\bibinfo {title} {Unified theory of
  dynamically suppressed qubit decoherence in thermal baths},\ }\href {\doibase
  10.1103/PhysRevLett.93.130406} {\bibfield  {journal} {\bibinfo  {journal}
  {Phys. Rev. Lett.}\ }\textbf {\bibinfo {volume} {93}},\ \bibinfo {pages}
  {130406} (\bibinfo {year} {2004})}\BibitemShut {NoStop}%
\bibitem [{\citenamefont {E.Slotine}\ and\ \citenamefont
  {Li}(1991)}]{Slotine1991}%
  \BibitemOpen
  \bibfield  {author} {\bibinfo {author} {\bibfnamefont {J.-J.}\ \bibnamefont
  {E.Slotine}}\ and\ \bibinfo {author} {\bibfnamefont {W.}~\bibnamefont {Li}},\
  }\href@noop {} {\emph {\bibinfo {title} {Applied nonlinear control}}}\
  (\bibinfo  {publisher} {Prentice Hall},\ \bibinfo {address} {Englewood
  Cliffs, New Jersey},\ \bibinfo {year} {1991})\BibitemShut {NoStop}%
\bibitem [{\citenamefont {Erez}\ \emph {et~al.}(2008)\citenamefont {Erez},
  \citenamefont {Gordon}, \citenamefont {Nest},\ and\ \citenamefont
  {Kurizki}}]{Erez2008}%
  \BibitemOpen
  \bibfield  {author} {\bibinfo {author} {\bibfnamefont {N.}~\bibnamefont
  {Erez}}, \bibinfo {author} {\bibfnamefont {G.}~\bibnamefont {Gordon}},
  \bibinfo {author} {\bibfnamefont {M.}~\bibnamefont {Nest}}, \ and\ \bibinfo
  {author} {\bibfnamefont {G.}~\bibnamefont {Kurizki}},\ }\bibfield  {title}
  {\bibinfo {title} {Thermodynamic control by frequent quantum measurements},\
  }\href {\doibase 10.1038/nature06873} {\bibfield  {journal} {\bibinfo
  {journal} {Nature}\ }\textbf {\bibinfo {volume} {452}},\ \bibinfo {pages}
  {724} (\bibinfo {year} {2008})}\BibitemShut {NoStop}%
\bibitem [{\citenamefont {Gordon}\ \emph {et~al.}(2009)\citenamefont {Gordon},
  \citenamefont {Bensky}, \citenamefont {Gelbwaser-Klimovsky}, \citenamefont
  {Rao}, \citenamefont {Erez},\ and\ \citenamefont {Kurizki}}]{Gordon2009}%
  \BibitemOpen
  \bibfield  {author} {\bibinfo {author} {\bibfnamefont {G.}~\bibnamefont
  {Gordon}}, \bibinfo {author} {\bibfnamefont {G.}~\bibnamefont {Bensky}},
  \bibinfo {author} {\bibfnamefont {D.}~\bibnamefont {Gelbwaser-Klimovsky}},
  \bibinfo {author} {\bibfnamefont {D.~D.~B.}\ \bibnamefont {Rao}}, \bibinfo
  {author} {\bibfnamefont {N.}~\bibnamefont {Erez}}, \ and\ \bibinfo {author}
  {\bibfnamefont {G.}~\bibnamefont {Kurizki}},\ }\bibfield  {title} {\bibinfo
  {title} {Cooling down quantum bits on ultrashort time scales},\ }\href
  {http://stacks.iop.org/1367-2630/11/i=12/a=123025} {\bibfield  {journal}
  {\bibinfo  {journal} {New J. Phys.}\ }\textbf {\bibinfo {volume} {11}},\
  \bibinfo {pages} {123025} (\bibinfo {year} {2009})}\BibitemShut {NoStop}%
\bibitem [{\citenamefont {Zhang}\ \emph {et~al.}(2019)\citenamefont {Zhang},
  \citenamefont {Jing}, \citenamefont {Wu}, \citenamefont {Wang},\ and\
  \citenamefont {Zhu}}]{Zhang2019}%
  \BibitemOpen
  \bibfield  {author} {\bibinfo {author} {\bibfnamefont {J.-M.}\ \bibnamefont
  {Zhang}}, \bibinfo {author} {\bibfnamefont {J.}~\bibnamefont {Jing}},
  \bibinfo {author} {\bibfnamefont {L.-A.}\ \bibnamefont {Wu}}, \bibinfo
  {author} {\bibfnamefont {L.-G.}\ \bibnamefont {Wang}}, \ and\ \bibinfo
  {author} {\bibfnamefont {S.-Y.}\ \bibnamefont {Zhu}},\ }\bibfield  {title}
  {\bibinfo {title} {Measurement-induced cooling of a qubit in structured
  environments},\ }\href {\doibase 10.1103/PhysRevA.100.022107} {\bibfield
  {journal} {\bibinfo  {journal} {Phys. Rev. A}\ }\textbf {\bibinfo {volume}
  {100}},\ \bibinfo {pages} {022107} (\bibinfo {year} {2019})}\BibitemShut
  {NoStop}%
\bibitem [{\citenamefont {Breuer}\ \emph {et~al.}(2001)\citenamefont {Breuer},
  \citenamefont {Kappler},\ and\ \citenamefont {Petruccione}}]{Breuer2001}%
  \BibitemOpen
  \bibfield  {author} {\bibinfo {author} {\bibfnamefont {H.-P.}\ \bibnamefont
  {Breuer}}, \bibinfo {author} {\bibfnamefont {B.}~\bibnamefont {Kappler}}, \
  and\ \bibinfo {author} {\bibfnamefont {F.}~\bibnamefont {Petruccione}},\
  }\bibfield  {title} {\bibinfo {title} {The time-convolutionless projection
  operator technique in the quantum theory of dissipation and decoherence},\
  }\href {\doibase https://doi.org/10.1006/aphy.2001.6152} {\bibfield
  {journal} {\bibinfo  {journal} {Ann. Phys.}\ }\textbf {\bibinfo {volume}
  {291}},\ \bibinfo {pages} {36} (\bibinfo {year} {2001})}\BibitemShut
  {NoStop}%
\bibitem [{\citenamefont {Weiss}(2008)}]{Weiss2008}%
  \BibitemOpen
  \bibfield  {author} {\bibinfo {author} {\bibfnamefont {U.}~\bibnamefont
  {Weiss}},\ }\href@noop {} {\emph {\bibinfo {title} {Quantum Dissipative
  Systems}}},\ \bibinfo {edition} {3rd}\ ed.\ (\bibinfo  {publisher} {Word
  Scientific},\ \bibinfo {year} {2008})\BibitemShut {NoStop}%
\bibitem [{\citenamefont {Baumgratz}\ \emph {et~al.}(2014)\citenamefont
  {Baumgratz}, \citenamefont {Cramer},\ and\ \citenamefont
  {Plenio}}]{Baumgratz2014}%
  \BibitemOpen
  \bibfield  {author} {\bibinfo {author} {\bibfnamefont {T.}~\bibnamefont
  {Baumgratz}}, \bibinfo {author} {\bibfnamefont {M.}~\bibnamefont {Cramer}}, \
  and\ \bibinfo {author} {\bibfnamefont {M.~B.}\ \bibnamefont {Plenio}},\
  }\bibfield  {title} {\bibinfo {title} {Quantifying coherence},\ }\href
  {\doibase 10.1103/PhysRevLett.113.140401} {\bibfield  {journal} {\bibinfo
  {journal} {Phys. Rev. Lett.}\ }\textbf {\bibinfo {volume} {113}},\ \bibinfo
  {pages} {140401} (\bibinfo {year} {2014})}\BibitemShut {NoStop}%
\bibitem [{\citenamefont {Forn-D\'{\i}az}\ \emph {et~al.}(2019)\citenamefont
  {Forn-D\'{\i}az}, \citenamefont {Lamata}, \citenamefont {Rico}, \citenamefont
  {Kono},\ and\ \citenamefont {Solano}}]{Diaz2019}%
  \BibitemOpen
  \bibfield  {author} {\bibinfo {author} {\bibfnamefont {P.}~\bibnamefont
  {Forn-D\'{\i}az}}, \bibinfo {author} {\bibfnamefont {L.}~\bibnamefont
  {Lamata}}, \bibinfo {author} {\bibfnamefont {E.}~\bibnamefont {Rico}},
  \bibinfo {author} {\bibfnamefont {J.}~\bibnamefont {Kono}}, \ and\ \bibinfo
  {author} {\bibfnamefont {E.}~\bibnamefont {Solano}},\ }\bibfield  {title}
  {\bibinfo {title} {Ultrastrong coupling regimes of light-matter
  interaction},\ }\href {\doibase 10.1103/RevModPhys.91.025005} {\bibfield
  {journal} {\bibinfo  {journal} {Rev. Mod. Phys.}\ }\textbf {\bibinfo {volume}
  {91}},\ \bibinfo {pages} {025005} (\bibinfo {year} {2019})}\BibitemShut
  {NoStop}%
\bibitem [{\citenamefont {Rivas}(2017)}]{Rivas2017}%
  \BibitemOpen
  \bibfield  {author} {\bibinfo {author} {\bibfnamefont {A.}~\bibnamefont
  {Rivas}},\ }\bibfield  {title} {\bibinfo {title} {Refined weak-coupling
  limit: Coherence, entanglement, and non-markovianity},\ }\href {\doibase
  10.1103/PhysRevA.95.042104} {\bibfield  {journal} {\bibinfo  {journal} {Phys.
  Rev. A}\ }\textbf {\bibinfo {volume} {95}},\ \bibinfo {pages} {042104}
  (\bibinfo {year} {2017})}\BibitemShut {NoStop}%
\bibitem [{\citenamefont {Jeske}\ \emph {et~al.}(2015)\citenamefont {Jeske},
  \citenamefont {Ing}, \citenamefont {Plenio}, \citenamefont {Huelga},\ and\
  \citenamefont {Cole}}]{Jeske2015}%
  \BibitemOpen
  \bibfield  {author} {\bibinfo {author} {\bibfnamefont {J.}~\bibnamefont
  {Jeske}}, \bibinfo {author} {\bibfnamefont {D.~J.}\ \bibnamefont {Ing}},
  \bibinfo {author} {\bibfnamefont {M.~B.}\ \bibnamefont {Plenio}}, \bibinfo
  {author} {\bibfnamefont {S.~F.}\ \bibnamefont {Huelga}}, \ and\ \bibinfo
  {author} {\bibfnamefont {J.~H.}\ \bibnamefont {Cole}},\ }\bibfield  {title}
  {\bibinfo {title} {Bloch-redfield equations for modeling light-harvesting
  complexes},\ }\href {\doibase 10.1063/1.4907370} {\bibfield  {journal}
  {\bibinfo  {journal} {J. Chem. Phys.}\ }\textbf {\bibinfo {volume} {142}},\
  \bibinfo {pages} {064104} (\bibinfo {year} {2015})}\BibitemShut {NoStop}%
\bibitem [{\citenamefont {Tscherbul}\ and\ \citenamefont
  {Brumer}(2015)}]{Tscherbul2015}%
  \BibitemOpen
  \bibfield  {author} {\bibinfo {author} {\bibfnamefont {T.~V.}\ \bibnamefont
  {Tscherbul}}\ and\ \bibinfo {author} {\bibfnamefont {P.}~\bibnamefont
  {Brumer}},\ }\bibfield  {title} {\bibinfo {title} {Partial secular
  bloch-redfield master equation for incoherent excitation of multilevel
  quantum systems},\ }\href {\doibase 10.1063/1.4908130} {\bibfield  {journal}
  {\bibinfo  {journal} {J. Chem. Phys.}\ }\textbf {\bibinfo {volume} {142}},\
  \bibinfo {pages} {104107} (\bibinfo {year} {2015})}\BibitemShut {NoStop}%
\bibitem [{\citenamefont {Majenz}\ \emph {et~al.}(2013)\citenamefont {Majenz},
  \citenamefont {Albash}, \citenamefont {Breuer},\ and\ \citenamefont
  {Lidar}}]{Majenz2013}%
  \BibitemOpen
  \bibfield  {author} {\bibinfo {author} {\bibfnamefont {C.}~\bibnamefont
  {Majenz}}, \bibinfo {author} {\bibfnamefont {T.}~\bibnamefont {Albash}},
  \bibinfo {author} {\bibfnamefont {H.-P.}\ \bibnamefont {Breuer}}, \ and\
  \bibinfo {author} {\bibfnamefont {D.~A.}\ \bibnamefont {Lidar}},\ }\bibfield
  {title} {\bibinfo {title} {Coarse graining can beat the rotating-wave
  approximation in quantum {Markovian} master equations},\ }\href {\doibase
  10.1103/PhysRevA.88.012103} {\bibfield  {journal} {\bibinfo  {journal} {Phys.
  Rev. A}\ }\textbf {\bibinfo {volume} {88}},\ \bibinfo {pages} {012103}
  (\bibinfo {year} {2013})}\BibitemShut {NoStop}%
\bibitem [{\citenamefont {Davidovi{\'{c}}}(2020)}]{Davidovic2020}%
  \BibitemOpen
  \bibfield  {author} {\bibinfo {author} {\bibfnamefont {D.}~\bibnamefont
  {Davidovi{\'{c}}}},\ }\bibfield  {title} {\bibinfo {title} {Completely
  {P}ositive, {S}imple, and {P}ossibly {H}ighly {A}ccurate {A}pproximation of
  the {R}edfield {E}quation},\ }\href {\doibase 10.22331/q-2020-09-21-326}
  {\bibfield  {journal} {\bibinfo  {journal} {{Quantum}}\ }\textbf {\bibinfo
  {volume} {4}},\ \bibinfo {pages} {326} (\bibinfo {year} {2020})}\BibitemShut
  {NoStop}%
\bibitem [{\citenamefont {Mozgunov}\ and\ \citenamefont
  {Lidar}(2020)}]{Mozgunov2020}%
  \BibitemOpen
  \bibfield  {author} {\bibinfo {author} {\bibfnamefont {E.}~\bibnamefont
  {Mozgunov}}\ and\ \bibinfo {author} {\bibfnamefont {D.}~\bibnamefont
  {Lidar}},\ }\bibfield  {title} {\bibinfo {title} {Completely positive master
  equation for arbitrary driving and small level spacing},\ }\href {\doibase
  10.22331/q-2020-02-06-227} {\bibfield  {journal} {\bibinfo  {journal}
  {{Quantum}}\ }\textbf {\bibinfo {volume} {4}},\ \bibinfo {pages} {227}
  (\bibinfo {year} {2020})}\BibitemShut {NoStop}%
\bibitem [{\citenamefont {Hartmann}\ and\ \citenamefont
  {Strunz}(2020)}]{Richard2020}%
  \BibitemOpen
  \bibfield  {author} {\bibinfo {author} {\bibfnamefont {R.}~\bibnamefont
  {Hartmann}}\ and\ \bibinfo {author} {\bibfnamefont {W.~T.}\ \bibnamefont
  {Strunz}},\ }\bibfield  {title} {\bibinfo {title} {Accuracy assessment of
  perturbative master equations: Embracing nonpositivity},\ }\href {\doibase
  10.1103/PhysRevA.101.012103} {\bibfield  {journal} {\bibinfo  {journal}
  {Phys. Rev. A}\ }\textbf {\bibinfo {volume} {101}},\ \bibinfo {pages}
  {012103} (\bibinfo {year} {2020})}\BibitemShut {NoStop}%
\bibitem [{\citenamefont {Gramich}\ \emph {et~al.}(2014)\citenamefont
  {Gramich}, \citenamefont {Gasparinetti}, \citenamefont {Solinas},\ and\
  \citenamefont {Ankerhold}}]{Gramich2014}%
  \BibitemOpen
  \bibfield  {author} {\bibinfo {author} {\bibfnamefont {V.}~\bibnamefont
  {Gramich}}, \bibinfo {author} {\bibfnamefont {S.}~\bibnamefont
  {Gasparinetti}}, \bibinfo {author} {\bibfnamefont {P.}~\bibnamefont
  {Solinas}}, \ and\ \bibinfo {author} {\bibfnamefont {J.}~\bibnamefont
  {Ankerhold}},\ }\bibfield  {title} {\bibinfo {title} {Lamb-shift enhancement
  and detection in strongly driven superconducting circuits},\ }\href {\doibase
  10.1103/PhysRevLett.113.027001} {\bibfield  {journal} {\bibinfo  {journal}
  {Phys. Rev. Lett.}\ }\textbf {\bibinfo {volume} {113}},\ \bibinfo {pages}
  {027001} (\bibinfo {year} {2014})}\BibitemShut {NoStop}%
\bibitem [{\citenamefont {Silveri}\ \emph {et~al.}(2019)\citenamefont
  {Silveri}, \citenamefont {Masuda}, \citenamefont {Sevriuk}, \citenamefont
  {Tan}, \citenamefont {Jenei}, \citenamefont {Hyyppa}, \citenamefont
  {Hassler}, \citenamefont {Partanen}, \citenamefont {Goetz}, \citenamefont
  {Lake}, \citenamefont {Gronberg},\ and\ \citenamefont
  {Mottonen}}]{Silveri2019}%
  \BibitemOpen
  \bibfield  {author} {\bibinfo {author} {\bibfnamefont {M.}~\bibnamefont
  {Silveri}}, \bibinfo {author} {\bibfnamefont {S.}~\bibnamefont {Masuda}},
  \bibinfo {author} {\bibfnamefont {V.}~\bibnamefont {Sevriuk}}, \bibinfo
  {author} {\bibfnamefont {K.~Y.}\ \bibnamefont {Tan}}, \bibinfo {author}
  {\bibfnamefont {M.}~\bibnamefont {Jenei}}, \bibinfo {author} {\bibfnamefont
  {E.}~\bibnamefont {Hyyppa}}, \bibinfo {author} {\bibfnamefont
  {F.}~\bibnamefont {Hassler}}, \bibinfo {author} {\bibfnamefont
  {M.}~\bibnamefont {Partanen}}, \bibinfo {author} {\bibfnamefont
  {J.}~\bibnamefont {Goetz}}, \bibinfo {author} {\bibfnamefont {R.~E.}\
  \bibnamefont {Lake}}, \bibinfo {author} {\bibfnamefont {L.}~\bibnamefont
  {Gronberg}}, \ and\ \bibinfo {author} {\bibfnamefont {M.}~\bibnamefont
  {Mottonen}},\ }\bibfield  {title} {\bibinfo {title} {Broadband lamb shift in
  an engineered quantum system},\ }\href {\doibase 10.1038/s41567-019-0449-0}
  {\bibfield  {journal} {\bibinfo  {journal} {Nat. Phys.}\ }\textbf {\bibinfo
  {volume} {15}},\ \bibinfo {pages} {533} (\bibinfo {year} {2019})}\BibitemShut
  {NoStop}%
\bibitem [{\citenamefont {Frey}\ \emph {et~al.}(2017)\citenamefont {Frey},
  \citenamefont {Mavadia}, \citenamefont {Norris}, \citenamefont {de~Ferranti},
  \citenamefont {Lucarelli}, \citenamefont {Viola},\ and\ \citenamefont
  {Biercuk}}]{Frey2017}%
  \BibitemOpen
  \bibfield  {author} {\bibinfo {author} {\bibfnamefont {V.~M.}\ \bibnamefont
  {Frey}}, \bibinfo {author} {\bibfnamefont {S.}~\bibnamefont {Mavadia}},
  \bibinfo {author} {\bibfnamefont {L.~M.}\ \bibnamefont {Norris}}, \bibinfo
  {author} {\bibfnamefont {W.}~\bibnamefont {de~Ferranti}}, \bibinfo {author}
  {\bibfnamefont {D.}~\bibnamefont {Lucarelli}}, \bibinfo {author}
  {\bibfnamefont {L.}~\bibnamefont {Viola}}, \ and\ \bibinfo {author}
  {\bibfnamefont {M.~J.}\ \bibnamefont {Biercuk}},\ }\bibfield  {title}
  {\bibinfo {title} {Application of optimal band-limited control protocols to
  quantum noise sensing},\ }\href {\doibase 10.1038/s41467-017-02298-2}
  {\bibfield  {journal} {\bibinfo  {journal} {Nat. Commun.}\ }\textbf {\bibinfo
  {volume} {8}},\ \bibinfo {pages} {2189} (\bibinfo {year} {2017})}\BibitemShut
  {NoStop}%
\bibitem [{\citenamefont {Norris}\ \emph {et~al.}(2018)\citenamefont {Norris},
  \citenamefont {Lucarelli}, \citenamefont {Frey}, \citenamefont {Mavadia},
  \citenamefont {Biercuk},\ and\ \citenamefont {Viola}}]{Norris2018}%
  \BibitemOpen
  \bibfield  {author} {\bibinfo {author} {\bibfnamefont {L.~M.}\ \bibnamefont
  {Norris}}, \bibinfo {author} {\bibfnamefont {D.}~\bibnamefont {Lucarelli}},
  \bibinfo {author} {\bibfnamefont {V.~M.}\ \bibnamefont {Frey}}, \bibinfo
  {author} {\bibfnamefont {S.}~\bibnamefont {Mavadia}}, \bibinfo {author}
  {\bibfnamefont {M.~J.}\ \bibnamefont {Biercuk}}, \ and\ \bibinfo {author}
  {\bibfnamefont {L.}~\bibnamefont {Viola}},\ }\bibfield  {title} {\bibinfo
  {title} {Optimally band-limited spectroscopy of control noise using a qubit
  sensor},\ }\href {\doibase 10.1103/PhysRevA.98.032315} {\bibfield  {journal}
  {\bibinfo  {journal} {Phys. Rev. A}\ }\textbf {\bibinfo {volume} {98}},\
  \bibinfo {pages} {032315} (\bibinfo {year} {2018})}\BibitemShut {NoStop}%
\bibitem [{\citenamefont {Bastidas}\ \emph {et~al.}(2012)\citenamefont
  {Bastidas}, \citenamefont {Emary}, \citenamefont {Regler},\ and\
  \citenamefont {Brandes}}]{Bastidas2012}%
  \BibitemOpen
  \bibfield  {author} {\bibinfo {author} {\bibfnamefont {V.~M.}\ \bibnamefont
  {Bastidas}}, \bibinfo {author} {\bibfnamefont {C.}~\bibnamefont {Emary}},
  \bibinfo {author} {\bibfnamefont {B.}~\bibnamefont {Regler}}, \ and\ \bibinfo
  {author} {\bibfnamefont {T.}~\bibnamefont {Brandes}},\ }\bibfield  {title}
  {\bibinfo {title} {Nonequilibrium quantum phase transitions in the dicke
  model},\ }\href {\doibase 10.1103/PhysRevLett.108.043003} {\bibfield
  {journal} {\bibinfo  {journal} {Phys. Rev. Lett.}\ }\textbf {\bibinfo
  {volume} {108}},\ \bibinfo {pages} {043003} (\bibinfo {year}
  {2012})}\BibitemShut {NoStop}%
\bibitem [{\citenamefont {De~Liberato}\ \emph {et~al.}(2009)\citenamefont
  {De~Liberato}, \citenamefont {Gerace}, \citenamefont {Carusotto},\ and\
  \citenamefont {Ciuti}}]{DeLiberato2009}%
  \BibitemOpen
  \bibfield  {author} {\bibinfo {author} {\bibfnamefont {S.}~\bibnamefont
  {De~Liberato}}, \bibinfo {author} {\bibfnamefont {D.}~\bibnamefont {Gerace}},
  \bibinfo {author} {\bibfnamefont {I.}~\bibnamefont {Carusotto}}, \ and\
  \bibinfo {author} {\bibfnamefont {C.}~\bibnamefont {Ciuti}},\ }\bibfield
  {title} {\bibinfo {title} {Extracavity quantum vacuum radiation from a single
  qubit},\ }\href {\doibase 10.1103/PhysRevA.80.053810} {\bibfield  {journal}
  {\bibinfo  {journal} {Phys. Rev. A}\ }\textbf {\bibinfo {volume} {80}},\
  \bibinfo {pages} {053810} (\bibinfo {year} {2009})}\BibitemShut {NoStop}%
\bibitem [{\citenamefont {Mori}\ and\ \citenamefont
  {Miyashita}(2008)}]{Mori2008}%
  \BibitemOpen
  \bibfield  {author} {\bibinfo {author} {\bibfnamefont {T.}~\bibnamefont
  {Mori}}\ and\ \bibinfo {author} {\bibfnamefont {S.}~\bibnamefont
  {Miyashita}},\ }\bibfield  {title} {\bibinfo {title} {Dynamics of the density
  matrix in contact with a thermal bath and the quantum master equation},\
  }\href {\doibase 10.1143/JPSJ.77.124005} {\bibfield  {journal} {\bibinfo
  {journal} {J. Phys. Soc. Jpn.}\ }\textbf {\bibinfo {volume} {77}},\ \bibinfo
  {pages} {124005} (\bibinfo {year} {2008})}\BibitemShut {NoStop}%
\bibitem [{\citenamefont {Trushechkin}\ \emph {et~al.}(2022)\citenamefont
  {Trushechkin}, \citenamefont {Merkli}, \citenamefont {Cresser},\ and\
  \citenamefont {Anders}}]{Trushechkin2022}%
  \BibitemOpen
  \bibfield  {author} {\bibinfo {author} {\bibfnamefont {A.~S.}\ \bibnamefont
  {Trushechkin}}, \bibinfo {author} {\bibfnamefont {M.}~\bibnamefont {Merkli}},
  \bibinfo {author} {\bibfnamefont {J.~D.}\ \bibnamefont {Cresser}}, \ and\
  \bibinfo {author} {\bibfnamefont {J.}~\bibnamefont {Anders}},\ }\bibfield
  {title} {\bibinfo {title} {Open quantum system dynamics and the mean force
  gibbs state},\ }\href {\doibase 10.1116/5.0073853} {\bibfield  {journal}
  {\bibinfo  {journal} {AVS Quantum Sci.}\ }\textbf {\bibinfo {volume} {4}},\
  \bibinfo {pages} {012301} (\bibinfo {year} {2022})}\BibitemShut {NoStop}%
\end{thebibliography}%

\end{document}